\documentclass[fleqn,twoside]{article}
\usepackage{espcrc2}
\usepackage{graphicx}
\usepackage[figuresright]{rotating}

\def\g2{{$(g-2)$}}
\newcommand{\AmS}{{\protect\the\textfont2
  A\kern-.1667em\lower.5ex\hbox{M}\kern-.125emS}}

\title{Muon $(g-2)$: Past, Present and Future }

\author{B. Lee Roberts\thanks{representing the muon $(g-2)$ 
    collaborations: E821 and E969.}
        \thanks{This work supported in part by the US National Science
Foundation and the U.S. Department of Energy.}\\
Department of Physics \\ 
        Boston University\\
        Boston, MA 02215}

\begin{document}

\begin{abstract}
The muon \g2 experiment E821 at the Brookhaven National Laboratory has
achieved a relative precision of $\pm 0.5$ parts per million. 
 A new experiment,
E969, with scientific approval but not yet funded, aims to improve this to
$\pm 0.2$~ppm.  The technique and results from E821 will be described, and the
proposed improvements for E969 will be discussed.
\vspace{1pc}
\end{abstract}

\maketitle

\section{Introduction}

A charged particle with spin $\vec s$ has a magnetic moment
$ \vec \mu_s = g_s ( {e / 2m} ) \vec s$; an anomaly
$a \equiv { (g_s -2) / 2}$; and  $ \mu = (1 + a){e \hbar / 2m}$;
where $g_s$ is the gyromagnetic ratio, 
and the latter expression is what one finds in the Particle
Data Tables.\cite{pdg}  The $g$-value is exactly 2 for a point-like
fermion in the Dirac equation, but radiative corrections give rise
to a non-zero value for the anomaly $a$.  The lowest order (QED) correction
gives $a = \alpha/2 \pi$.  For the muon,
radiative corrections from QED, virtual hadrons (quarks), and weak gauge
bosons are important at the level of measurement.\cite{davmar}
While the QED and weak contributions can be calculated to the necessary
accuracy to compare with experiment, the hadronic contribution needs to
be obtained using data from  $e^+ e^- \rightarrow {\rm hadrons}$.

In a series of three experiments at CERN the muon anomaly was measured
to a relative precision of 7.3 parts per million (ppm).\cite{cern}
 Experiment E821
at the Brookhaven Alternating Gradient Synchrotron improved on this by 
a factor of 14,\cite{carey,brown1,brown2,bennett1,bennett2} to a 
relative precision of 0.5~ppm.  

While the value of the hadronic 
contribution has changed  over time, the other contributions have
remained quite steady.\cite{davmar}  
During all of this, there has remained a 
discrepancy between theory and
experiment of between two and three standard deviations
when the hadronic contribution is taken from 
$e^+e^-$ data.  When hadronic $\tau$-decay and CVC theory
is used to determine the hadronic contribution, the discrepancy is
smaller, but there appear to be as yet not understood isospin violation
corrections which make it difficult to compare the two methods.\cite{davmar}
Theoretically, the $e^+e^-$ cross-section is what enters into the 
dispersion integral.
Motivated by this potential discrepancy, a new experiment, E969 has 
been proposed at Brookhaven to improve the precision to 0.2~ppm.

\section{Measurement of the muon anomaly}
The method used in the third
CERN experiment and the BNL experiment are
very similar, save the use of superconducting magnets
for the storage ring and inflector, as well as direct muon 
injection into the storage ring.
These experiments are based on the
fact that for  $a_{\mu} > 0$ the spin 
precesses faster than
the momentum vector when a muon travels transversely to a 
magnetic field.  
The difference frequency between the cyclotron frequency and the
muon spin precession frequency,
$\omega_a = \omega_S - \omega_C = \left(( g-2) / 2 \right) ({eB / mc})$
is the frequency with which the spin
precesses relative to the momentum, and is  proportional to
the anomaly, rather than to $g$.
With both an electric
and a magnetic field, the spin difference frequency is given by
\begin{equation}
\vec \omega_a = - {e\over mc}
\left[ a_{\mu} \vec B -
\left( a_{\mu}- {1 \over \gamma^2 - 1}\right) \vec \beta 
\times {\vec E }
\right],
\label{eq:tbmt}
\end{equation}
which reduces to the simpler equation above 
in the absence of an electric field.
Electric quadrupoles were used for vertical focusing, 
taking advantage of the 
``magic''~$\gamma=29.3$ at which an electric field does not contribute to
the spin motion relative to the momentum. 

A precision measurement of $a_{\mu}$ requires precision measurements
of the muon spin precession frequency $\omega_a$,  and the magnetic field,
which is expressed as the free-proton precession frequency
$\omega_p$ in the storage ring magnetic field.
These two (average) frequencies plus the fundamental 
constant $\lambda = \mu_{\mu}/\mu_p$ give the anomaly:
\begin{equation}
a_{\mu} = { {\omega_a / \omega_p} \over {\lambda - \omega_a / \omega_p}} .
\end{equation}

The experimental signal is the $e^{\pm}$ from $\mu^{\pm}$ decay, 
detected by lead-scintillating
fiber calorimeters.   Since the highest
energy  $e^{\pm}$ are correlated with the muon spin, if one counts high-energy 
 $e^{\pm}$ as a function of time, one gets an exponential from muon decay
modulated by the $(g-2)$ precession. The expected form for the positron time
spectrum is $f(t) =  {N_0} e^{- \lambda t } 
[ 1 + {A} \cos ({\omega_a} t + {\phi})] $, however in analyzing the
data it is  necessary
to take a number of small effects into account.\cite{bennett1,bennett2}

\begin{figure}[h!]
  \includegraphics[width=.35\textwidth]{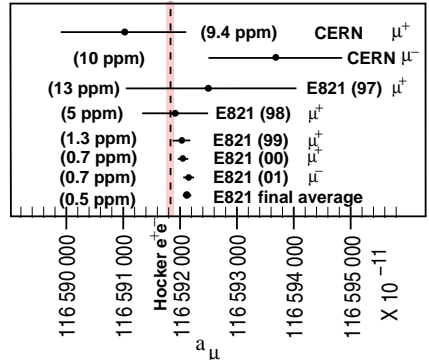}
  \caption{Measurements of $a_{\mu}$. The strong interaction 
contribution is taken from 
reference \cite{HMNTHICHEP}. }
\label{fg:amu}
\end{figure}

The values obtained for $a_{\mu}$ by E821 are shown in Fig.~\ref{fg:amu}
along with one theory value using $e^+e^-$ data for the lowest-order hadronic
contribution.\cite{HMNTHICHEP}  The discrepancy with theory
varies between 2.2 and 2.8 standard deviations when using the $e^+e^-$ 
data for the hadronic contribution, and about one-third of this when 
using the $\tau$-data.  The improvement of the $e^+e^-$ data, and the
understanding of the related theoretical issues is under active study
worldwide.\cite{eps05}

\section{Future Muon \g2 Experiments}

To increase the sensitivity of the comparison between theory and 
experiment, a new collaboration was
formed to continue the measurement of \g2 at BNL.  The goal is a total
error of 0.2~ppm, a factor of 2.5 better than E821.  This increased
precision,
combined with the expected improvements in the knowledge of the hadronic
contribution should give at least a factor of two reduction in the combined 
experiment-theory uncertainty when comparing theory with experiment.
In September 2004 the new experiment E969\cite{E969}
received enthusiastic scientific endorsement by the Laboratory.
The funding situation is less clear, and E9669 is expected to be considered 
by the HEPAP sub-panel P5 in early 2006.  

E821 achieved a final uncertainty on the
measurement of the muon anomalous magnetic moment $a_{\mu}$ of
0.54~ppm, which is dominated by the statistical error
of 0.46~ppm.  For our last data set the systematic uncertainties 
on the knowledge of
$\langle B \rangle$ and $\omega_a$ were 0.17~ppm and 0.21~ppm respectively,
for a total systematic of 0.27~ppm.

A further increase in precision
is possible if a higher muon storage rate can be obtained,
and the systematic uncertainties present in E821 are reduced. 
The proposed 0.2~ppm uncertainty is derived
from a 0.14~ppm statistical error, and equal total systematic
uncertainties of 0.1~ppm from the measurements of $\omega_p$
and $\omega_a$.  Ten times
more events compared to E821 are needed.

\begin{figure}
  \includegraphics[width=0.45\textwidth,angle=0]{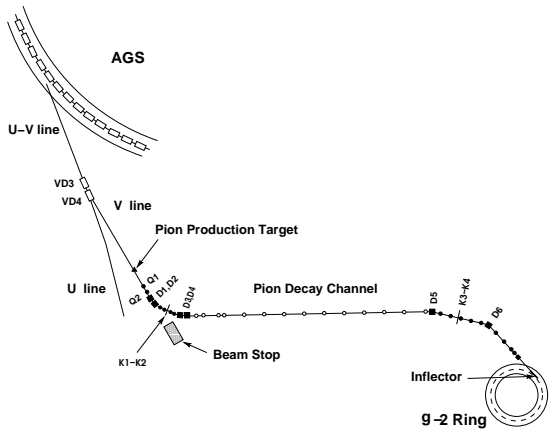}
  \caption{The E821 Beamline.  Pions produced at $0^{\circ}$
are collected in Q1Q2 and the momentum is selected on collimators
K1-K2.  The pion decay channel is 80 m in length.  Forward muons
are selected at the collimator K3K4.
  \label{fg:bmline}}
\end{figure}

An important feature of the upgraded experiment is a new front-end to
the beamline.  In E821 pions slightly higher than the magic momentum 
decay in a FODO decay channel, as shown in Fig.~\ref{fg:bmline}.
Forward decay $\mu$
were accepted at a momentum slit, but because of the large tail 
on the $\pi$ distribution, the $\pi:\mu$ ratio was about 1:1.  In
the new experiment backward decays in a $\pi$ beam of 5.3 GeV/c 
will be used to produce muons of 3.1 GeV, thus eliminating the baseline
shifts in the detectors that were caused by the pion ``flash'' at injection.
To increase the muon flux, we will double the number of quadrupoles in the
decay channel, and open the ends of the inflector magnet.\cite{infl}
This will gain a factor of four in muon flux. 

A number of experimental systems will need to be upgraded for E969.
To handle the increased rates new segmented detectors and their downstream
electronics will be developed.  The magnetic field measurement and
control will need to be improved, and the magnet will be shimmed
further.  Also, the muon kicker will need to be
upgraded.  All of these issues are presently under study.

While it is possible to improve the experimental value of $a_{\mu}$ further,
to below
0.1~ppm,  the motiviation for such an improvement would need to be 
driven by a better understanding of the hadronic contribution. Certainly
our upgraded experiment E969, combined with expected improvements in the
knowledge of the hadronic contribution, will present the community with
a new, more stringent test of the standard model.  It is clear that interest
in our result, and in our ability to improve upon it, remains high.

\end{document}